\documentclass[%
 reprint,
superscriptaddress,
 amsmath,amssymb,
 prl,
]{revtex4-2}

\usepackage{graphicx}
\usepackage[export]{adjustbox}
\usepackage{dcolumn}
\usepackage{bm}

\usepackage{braket}
\usepackage{graphicx}
\usepackage{amsmath}
\usepackage{mathrsfs}
\usepackage[dvipsnames]{xcolor}
\usepackage{longtable}
\usepackage{amsfonts}
\usepackage{tikz}
\usepackage[colorlinks,linkcolor=blue,citecolor=blue,anchorcolor=blue]{hyperref}
\usepackage{CJK}
\usepackage{upgreek}
\usepackage{tikz}
\usetikzlibrary{quantikz}
\usepackage{xr}
\usepackage{float}
\usepackage{siunitx}
\usepackage{multirow}
\usepackage{amsmath}
\usepackage[normalem]{ulem}
\usepackage[T1]{fontenc}

\begin{document}

\preprint{APS/123-QED}

\title{ 
Site-selective cavity readout and classical error correction of a 5-bit atomic register
}

\author{Beili Hu}
\thanks{These authors contributed equally to this work.}
\author{Josiah Sinclair}
\thanks{These authors contributed equally to this work.}
\author{Edita Bytyqi}
\author{Michelle Chong}
\author{Alyssa Rudelis}
\author{Joshua Ramette}
\author{Zachary Vendeiro}
\author{Vladan Vuleti\'c}
\thanks{vuletic@mit.edu}
\affiliation{%
 Department of Physics, MIT-Harvard Center for Ultracold Atoms and Research Laboratory of Electronics, Massachusetts Institute of Technology, Cambridge, Massachusetts 02139, USA\\
 }%

\date{\today} 

\begin{abstract}
Optical cavities can provide fast and non-destructive readout of individual atomic qubits; however, scaling up to many qubits remains a challenge.
Using locally addressed excited-state Stark shifts to tune atoms out of resonance, we realize site-selective hyperfine-state cavity readout across a 10-site array. 
The state discrimination fidelity is 0.994(1) for one atom and 0.989(2) averaged over the entire array at a survival probability of 0.975(1).
To further speed up array readout, we demonstrate adaptive search strategies utilizing global/subset checks.
Finally, we demonstrate repeated rounds of classical error correction, showing exponential suppression of logical error and extending logical memory fivefold beyond the single-bit idling lifetime.
\end{abstract}

\maketitle

Neutral-atom arrays are rapidly advancing toward large-scale quantum error correction (QEC).
Notable recent demonstrations include 6000-qubit arrays \cite{manetsch2024}, 
sub-microsecond one- and two-qubit gates with fidelities surpassing QEC thresholds \cite{Levine2022, Ma2023, Evered2023, tsai2024, radnaev2024}, coherent transport of entangled qubits \cite{Bluvstein2022},
and operations on 48 logical qubits showing reduction of the logical error rate with code distance \cite{Bluvstein2024}.
However, neutral-atom QEC experiments to date have been limited to a single round of error correction due to destructive qubit readout \cite{Bluvstein2024}. 
To realize repeated error correction, parallel non-destructive qubit readout \cite{Kwon2017, MartinezDorantes2017, Wu2019}
and continuous reloading \cite{gyger2024, Norcia2024} are being explored.
However, the timescales in these demonstrations currently lag behind other operational timescales, limiting logical clock speeds.

Cavity-based state detection is a promising alternative approach for fast and non-destructive readout. 
Through the Purcell effect \cite{Purcell1946}, a cavity enhances atomic emission into the cavity mode over emission into free space, increasing collection efficiency and enabling atomic-state detection after scattering only a few photons. 
Making use of this enhancement, multiple experiments have coupled a single atom to an optical cavity and demonstrated fast (tens of microseconds), low-loss ($<1\%$) and high-fidelity ($>99\%$) state detection \cite{Bochmann2010, Gehr2010, Deist2022}.
Scaling up cavity-enhanced readout to many atoms requires interfacing an atom array with a cavity, which has only been recently addressed \cite{Dordevic2021, Yan2023, Liu2024, Hartung2024}.
As an additional challenge to scaling, the cavity couples globally to all atoms inside its mode volume, prohibiting site-resolved readout of multiple atoms.
Proposed strategies to scale up cavity readout include shelving atoms in a dark state \cite{Kwon2017}, shifting one qubit level out of resonance \cite{Urech2022}, or assembling an array of cavities \cite{Trupke2007, Derntl2014, Wachter2019, shadmany2024}.
Very recently, cavity measurement of one atom with no observable hyperfine-state decoherence on a second atom outside the cavity mode has been demonstrated in combination with atom transport in $\SI{200}{\mu s}$~\cite{Deist2022}. 
However, measuring multiple qubits sequentially potentially forfeits any speedup compared to parallel free-space imaging methods. 

\begin{figure}[h]
\includegraphics[width=8.6cm]{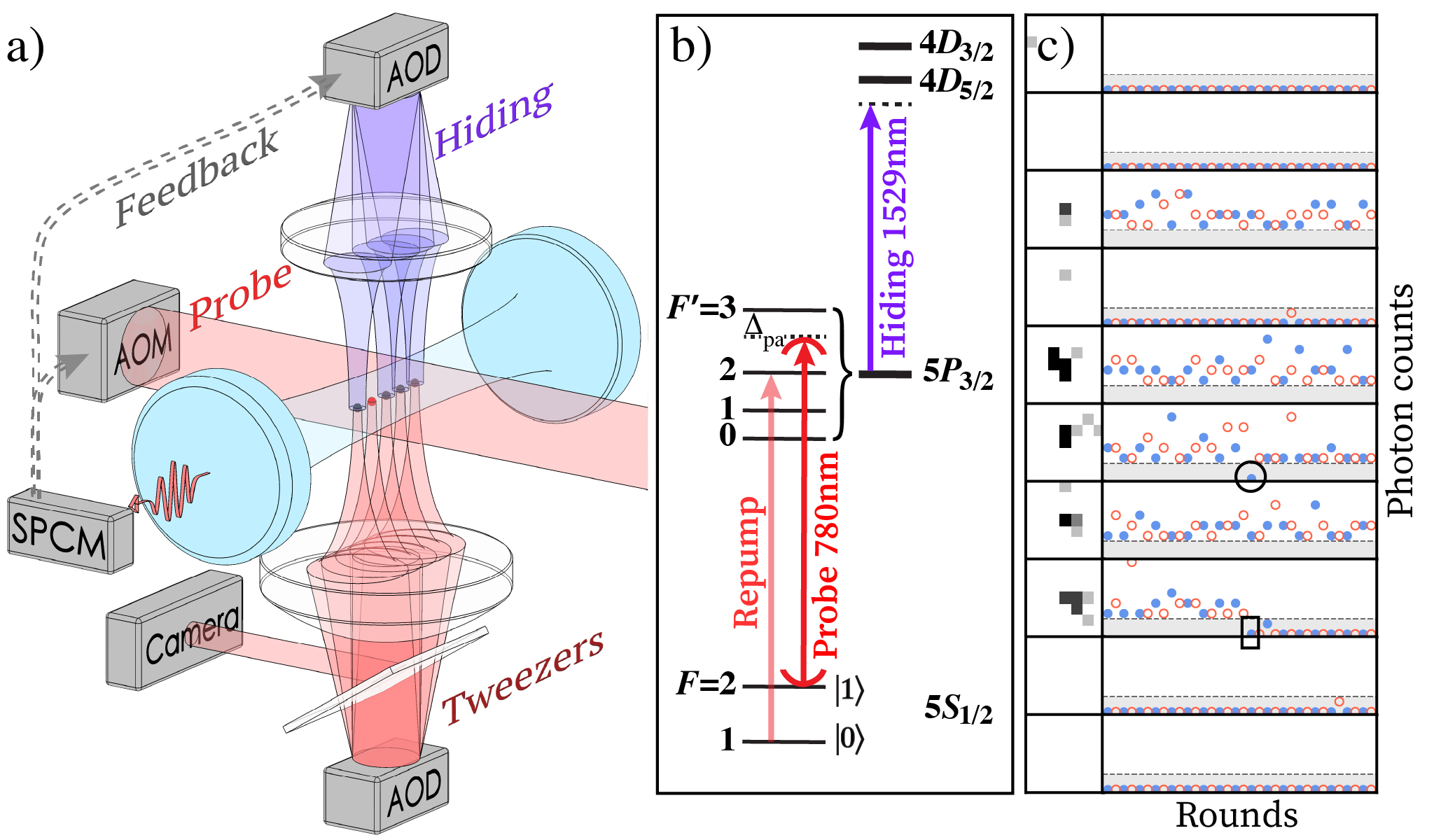}
\caption{
(a) Experimental setup.
Individual atoms are loaded into a one-dimensional tweezer array along the cavity axis, illuminated by a global probe beam and individually addressed hiding beams. 
Fluorescence is collected from both sides of the cavity (only one side shown) into single-photon counting modules (SPCMs). 
The probe beam and the hiding beams are controlled adaptively in real time by measurement results on the SPCMs. 
(b) 
$^{87}$Rb level diagram. 
(c) 
Example experiment with 10 tweezers and sequential adaptive readout. 
Left: camera image indicating tweezer occupation.
Right: collected photon counts from repeated rounds of state preparation in $F=2$ and adaptive cavity readout.
In each round, ten tweezers are probed, alternating between hyperfine (blue solid symbols) and occupation (pink open symbols) measurements.
Grey shaded area indicates detection threshold; black circle (rectangle) indicates a bit flip error (atom loss). 
}
\label{fig1}
\end{figure}

In this Letter, we implement a ``hiding'' strategy for scaling cavity readout to an array of atoms.  
By locally light-shifting individual atoms on an excited-to-excited-state transition \cite{Clark2020, Browaeys2022, Thompson2022, Norcia2023, Browaeys2024, Barnes2022}, we can address any subset of the array on a fast timescale, with minimal effect on the hyperfine ground states.
To demonstrate our approach, we sequentially read out an array containing up to 10 atoms and show that hidden atoms are well-shielded against probe-induced depumping with only milliwatts of hiding power per atom. 
To further speed up readout tasks such as syndrome readout for quantum error correction, we implement an ``adaptive" search strategy utilizing real-time decision-making, and observe a substantial improvement over deterministic sequential readout.
Finally, previewing how cavity readout with real-time feedback can enable repeated rounds of error correction, we implement a classical repetition code, where one bit of information is redundantly encoded in the hyperfine states of an atomic register. 
We observe exponentially suppressed errors for $d=3$ and $5$ logical bits, and extend the logical bit lifetime fivefold beyond the single-bit idling lifetime, limited mainly by percent-level atom loss.

In our experiment (see Fig.\:\ref{fig1}), $^{87}$Rb atoms are laser-cooled and stochastically loaded into a one-dimensional array of 10 tweezers, formed by \SI{808}{nm} beams generated by an acousto-optical deflector (AOD) and focused through a 0.55 numerical aperture (NA) in-vacuum aspheric lens. The tweezers are located along the axis of an in-vacuum standing-wave optical cavity with a mirror spacing of \SI{4.39}{cm} and a waist of $\SI{45}{\micro\meter}$.
The cavity finesse is $\mathcal{F}=34,000$, giving a maximum cooperativity of $\eta_0=4g_0^{2}/\kappa \Gamma = 2$, with $\{2g_{0}, \kappa, \Gamma\} = 2\pi \times \{1.1, 0.10, 6.0\}$ MHz, where $2g_{0}$ is the maximum single-photon Rabi frequency on the $^{87}$Rb $D_2$ cycling transition, and $\kappa$ and $\Gamma$ are the decay rates of the cavity and of the atomic excited state, respectively \cite{Rudelis2023}.
When the atoms are laser-cooled inside the optical tweezers, light-assisted collisions reduce the population of each tweezer site to either zero or one atom \cite{Schlosser2001}, and the resulting fluorescence is collected through the aspheric lens and imaged on a CMOS camera. 
To encode quantum information in the ground state hyperfine manifolds, we optically pump the atoms with a global depump ($F=2\rightarrow F'=2$) or repump ($F=1\rightarrow F'=2$) pulse. 
Thus, each tweezer in the array is either unoccupied or occupied with an atom prepared in the $F=1$ or $F=2$ ground state manifold. 

We first characterize cavity readout for a single tweezer.
The tweezer is illuminated from the side with a pair of counter-propagating probe beams, each with waist \SI{0.9}{mm} and power \SI{36}{\micro W}, detuned by $\Delta_{pa}/(2\pi) = -\SI{5}{MHz}$ below the $D_2$ $F=2\rightarrow F'=3$ transition and in a lin$\perp$lin configuration for polarization gradient cooling. 
The cavity resonance frequency is tuned to be $\kappa/2$ above the probe frequency for cavity cooling \cite{VuleticPRA2001} of the atom. 
The resulting fluorescence is collected from both sides of the cavity onto two single-photon counting modules (SPCM), with a total quantum efficiency $QE=0.27$.
We use a pair of $\SI{200}{\micro s}$-long cavity-enhanced fluorescence measurements to determine tweezer occupation and atomic hyperfine state.
In the first measurement interval, the number of collected photons indicates the hyperfine state of the atom, with $F=2$ being the bright state and $F=1$ the dark state. 
In the second interval, we detect atom presence by turning on a repumper beam.
The atom is then re-cooled and optically pumped before the next measurement. 

\begin{figure}
\includegraphics[width=8.6cm]{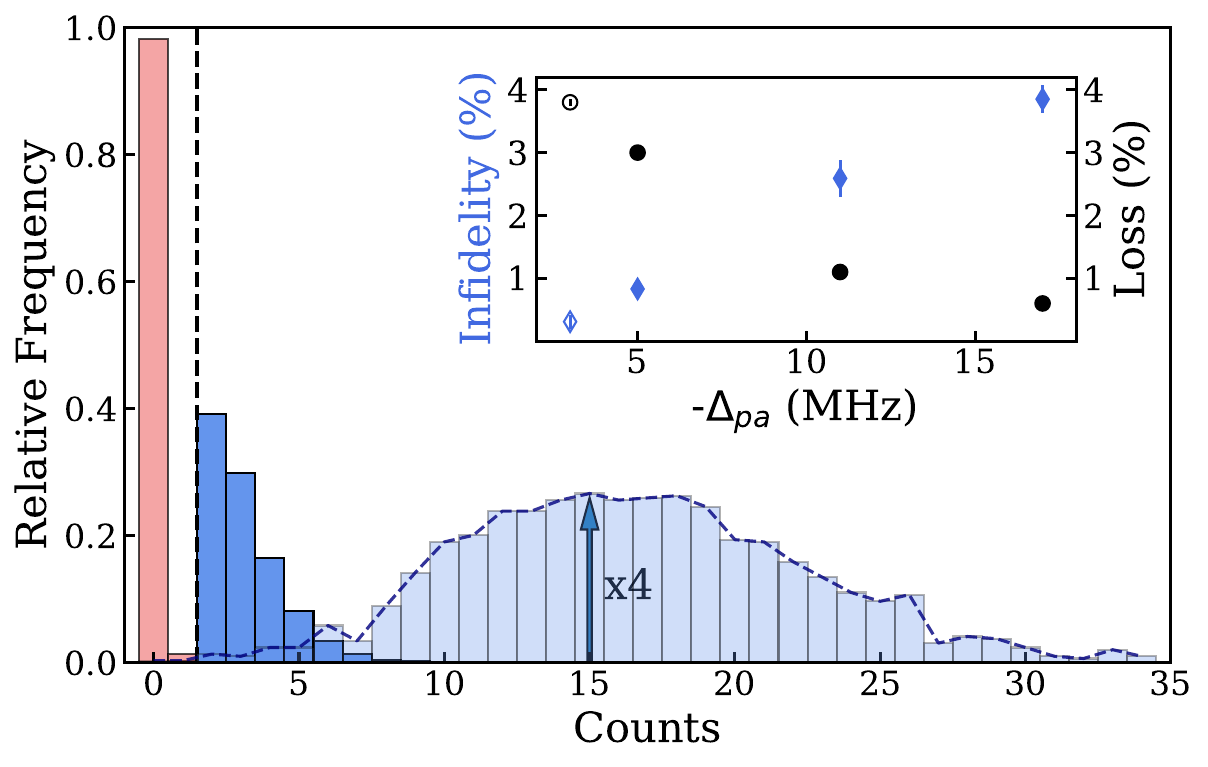}
\caption{Hyperfine-state readout ($F$=1, 2) of a single atom via photon scattering into the cavity. 
Light pink: $F$=1; light blue: $F$=2 full measurement interval; dark blue: $F$=2 adaptive measurement. 
Inset: observed trade-off between atom loss (black circles) and state infidelity (blue diamonds) for an $F=2$ atom for tweezer depth $U/k_B=0.25$~mK (solid symbols) and $U/k_B=0.2$~mK (open symbols).
}
\label{single_atom}
\end{figure}

Fig. \ref{single_atom} shows a histogram of collected photon counts in the first interval, conditioned on a positive atomic-presence detection in the second interval.
The detection threshold is drawn between 1 and 2 photons, limited by the $\SI{60}{s^{-1}}$ dark count rate of the SPCMs.
Over a full $\SI{200}{\micro s}$ measurement, we collect an average of 15 photons from an $F=2$ atom (light blue), and much less than 1 photon from an $F=1$ atom (light pink). 
However, photons that arrive after the threshold has been surpassed provide no additional information and only contribute to atom heating resulting in loss. 
To suppress this measurement-induced loss, we use a Quantum Machines OPX device to implement an adaptive measurement in all subsequent results, checking the SPCM counts every 20~$\mu s$ and switching off the probe beams as soon as the number of detected photons surpasses the detection threshold \cite{Hume2007, Myerson2008, Gibbons2011, Todaro2021, Erickson2022, Chow2023}. 
The resulting photon number distribution (dark blue) is highly non-Poissonian, rising sharply just past the detection threshold.
The tail of the distribution comes from multiple photons arriving in the same sub-interval.
Terminating the measurement in real time reduces the average number of photons collected from the bright state by a factor of $5.3(1)$, and reduces bright-state atom loss by a factor of $4.5(2)$.

\begin{table}[h!]
\centering
\begin{tabular}{|c|c|cc|cc|}
\hline
\multirow{2}{*}{$U/k_\textrm{B}$} & \multicolumn{1}{l|}{\multirow{2}{*}{$-\Delta_{pc}/2\pi$}} & \multicolumn{2}{c|}{$F=1$}               & \multicolumn{2}{c|}{$F=2$}               \\ \cline{3-6} 
                                       & \multicolumn{1}{l|}{}                                     & \multicolumn{1}{c|}{Infidelity} & Loss   & \multicolumn{1}{c|}{Infidelity} & Loss   \\ \hline
\SI{0.20}{mK}                          & 3~MHz                                                     & \multicolumn{1}{c|}{0.17(7)}    & 3.0(1) & \multicolumn{1}{c|}{0.3(1)}     & 3.8(1) \\ \hline
\SI{0.25}{mK}                           & 5~MHz                                                     & \multicolumn{1}{c|}{0.39(7)}    & 2.1(1) & \multicolumn{1}{c|}{0.8(1)}     & 3.0(1) \\ \hline
\SI{0.25}{mK}                           & 11~MHz                                                    & \multicolumn{1}{c|}{0.30(8)}    & 0.7(1) & \multicolumn{1}{c|}{2.6(3)}     & 1.1(1) \\ \hline
\SI{0.25}{mK}                           & 17~MHz                                                    & \multicolumn{1}{c|}{0.36(5)}    & 0.3(1) & \multicolumn{1}{c|}{3.9(2)}     & 0.6(1) \\ \hline
\end{tabular}
\caption{State detection infidelity and loss probability (both in \%) for cavity readout of a single atom, performed at different tweezer depths $U$ and probe-cavity detunings $-\Delta_{pc}$.}
\label{single_atom_table}
\end{table}

The state-averaged measurement fidelity is $99.4(1)\%$ (see Table \ref{single_atom_table}), and is limited by the moderate cavity cooperativity and \SI{5}{MHz} detuning from the cycling transition, effectively increasing the probability for off-resonant hyperfine-changing scattering events into free space.
The measured scattering rate and fidelity agree within 50\% with the result of a Jaynes-Cummings Hamiltonian simulation \cite{JOHANSSON20131234}, taking into account the spatially averaged cooperativity over nodes and anti-nodes, a threefold reduction in cavity scattering rate due to mixed probe polarizations \cite{Deist2022}, as well as cavity filtering of a Doppler-broadened emission spectrum. 
Our cavity readout supports a trade-off between loss and bright-state hyperfine fidelity (inset of Fig.\:\ref{single_atom} and Table\:\ref{single_atom_table}). 
Choosing a smaller probe-atom detuning suppresses off-resonant, hyperfine-changing scattering events improving measurement fidelity. 
However, polarization gradient cooling becomes less effective, resulting in more loss.
Similarly, with a shallower tweezer, an atom is ejected after fewer measurements, but the lower atomic temperature also gives rise to a narrower Doppler emission spectrum, improving the effective cavity cooperativity and hence the fidelity.

We next demonstrate scalable site-selective readout of an atom array coupled to an optical cavity. 
By applying individually addressed light shifts to excited states, we hide all but the target atom from the probe beam, suppressing scattering and protecting hyperfine information while performing a cavity readout of the target atom. 
At \SI{1529.420}{nm}, the light-shifting beam is \SI{7}{GHz} below the $5P_{3/2} \rightarrow 4D_{5/2}$ transition and \SI{20}{GHz} below the weaker $5P_{3/2} \rightarrow 4D_{3/2}$ transition, resulting in a polarizability that is about 22,000 times larger for the $5P_{3/2}$ excited state than for the $5S_{1/2}$ ground state \cite{Sibalic2017}. 
This disparity allows us to rapidly toggle $\SI{2}{GHz}$ light shifts on the excited state, without jostling atoms in the ground state.
Moreover, by choosing to shift non-target atoms out of resonance, rather than target atoms into resonance, the system is robust to alignment and power drifts. 

The light-shifting beams are generated by an AOD, allowing fast switching on the order of a few $\mu$s.
By dragging a single atom across a beam and observing the change in scattering rate, we measure a beam waist of $\SI{4}{\mu m}$ and a light shift of $\SI{1}{MHz/\mu W}$ at the beam center in the low-intensity limit \cite{Deist2022_superresolution}.
We observe 1\% residual light-shift at around $\SI{10}{\mu m}$ from the beam center, which we attribute largely to optical aberrations arising from focusing through an aspheric lens at a non-design wavelength.
To avoid cross-talk, atoms are spaced $\SI{17}{\mu m}$ apart in all measurements, limiting the array size, but closer spacing of atoms should be possible with an improved optical system. 

\begin{figure}[h]
\includegraphics[width=8.6cm]{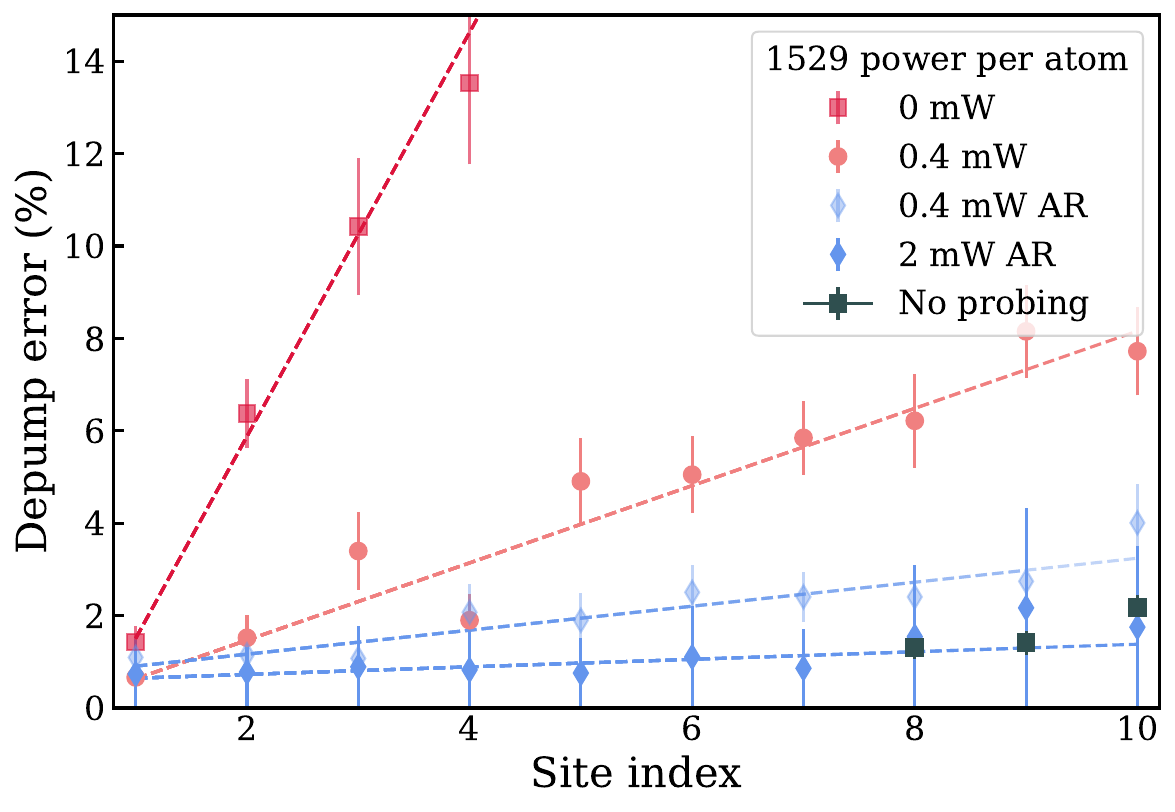}
\caption{Bright-state depump errors for a sequential cavity readout of up to 10 atoms prepared in $F=2$. 
AR: adaptive rounds, where in every round of sequential measurements, we only measure atoms present in the last round. 
No probing: depump errors from waiting in the tweezer for 7, 8, and 9 intervals, with no probe applied.  
\SI{0}{mW}: the same single atom is measured repeatedly to characterize probe-induced depump errors.
For all measurements, a tweezer depth of $U/k_\textrm{B} =$ \SI{0.25}{mK} and a probe detuning of \SI{5}{MHz} were used.
For the highest hiding power used, the $y$-intercept is fitted to $0.6(1)\%$ and the slope to $0.08(3)\%$/site, corresponding to the single-atom SPAM error and the additional error from scaling, respectively, the latter being dominated by depumping due to idling in \SI{808}{nm} tweezers.
}
\label{multi_atom}
\end{figure}

Figure \ref{multi_atom} shows the state preparation and measurement (SPAM) errors in a sequential measurement of 10 sites, using a tweezer depth of $U/k_\textrm{B} =$ \SI{0.25}{mK}, a probe detuning of \SI{5}{MHz}, and all initialized in the bright state $F=2$.
To calibrate how much the hiding beams suppress probe-induced depump errors ($F=2 \rightarrow F=1$) in a sequential measurement, we first separately characterize the depump errors by loading and repeatedly measuring a single atom, finding an increase of $4.4(3)\%$ at the end of each full probe interval. 
With an intermediate hiding power of \SI{0.4}{mW} per atom, depump errors are suppressed by a factor of $5.2(6)$.
Depump errors can be even further suppressed using an adaptive strategy where we only measure tweezers which were occupied in the previous round.
Combining this adaptive strategy with \SI{2}{mW} of hiding power per atom suppresses the probe-induced depumping sufficiently to be indistinguishable from background depumping from the \SI{808}{nm} optical tweezer, at $0.08(3)\%$ per interval.

When scaling to larger arrays, a sequential readout erodes the cavity's speed advantage. 
However, if atoms are biased to be in the dark state, faster-than-sequential strategies are available.
In this case, it becomes advantageous to unhide and read out groups of atoms at a time, using a single fluorescence interval to reveal if any bright state atom(s) are present. 
If not, the readout is complete; otherwise, locating the error then resembles a search problem, and can be sped up by further partitioning the array. 
This scenario is relevant to quantum error correction, where all syndrome qubits will be highly biased in a code operating below threshold (typically $1\%$). 

\begin{figure}
\includegraphics[width=8.6cm]{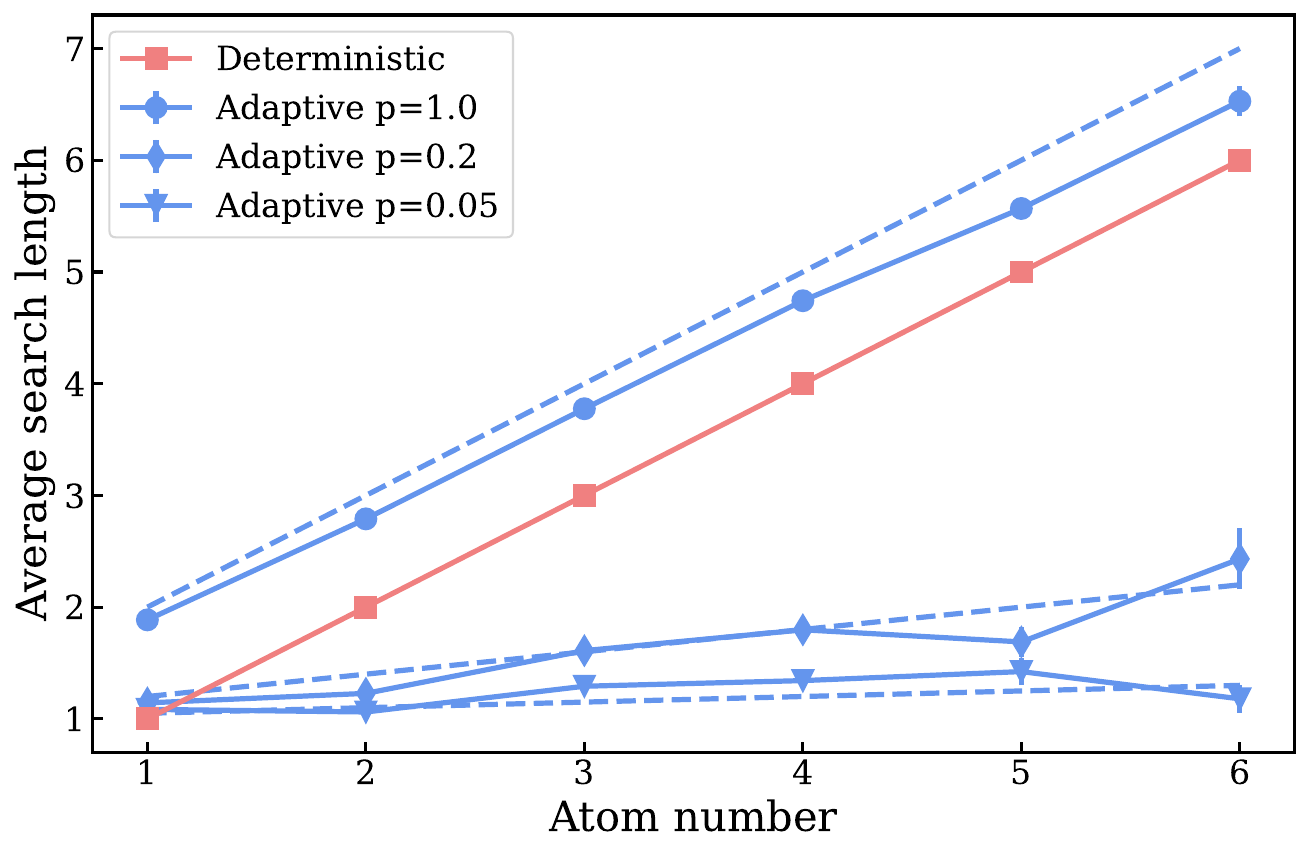}
\caption{Comparing the average length of deterministic and adaptive search strategies. The experimentally implemented adaptive strategy (blue) consists of a sequential search conditioned on the result of a global check. The length of an adaptive search depends linearly on atom number and the probability $p$ that a bright atom is present in the array. Dashed lines depict $1 + pN$. For a deterministic search (red squares) each atom would be checked sequentially, scaling as $N$. 
}
\label{adaptive_search}
\end{figure}

To demonstrate speed-up in reading out biased qubits, we prepare a register of N atoms in the dark state $F=1$, and with probability $p$ pump one atom into the bright state $F=2$.
For our small system size and a low error probability, the main speed-up comes from doing an initial global check that often does not find any bright-state atom, concluding the register readout in a single interval.
We therefore implement a simple adaptive strategy:
in the event where the global check detects a bright-state atom, we locate it by sequentially reading out the entire register.
As shown in Fig.\:\ref{adaptive_search}, this reduces the average number of cavity readout intervals from $N$ to $1+pN$. 
For larger system sizes, the search can be further sped up to $1+p\textrm{log}_2N$ by partitioning the array.

\begin{figure*}
\includegraphics[width=\textwidth]{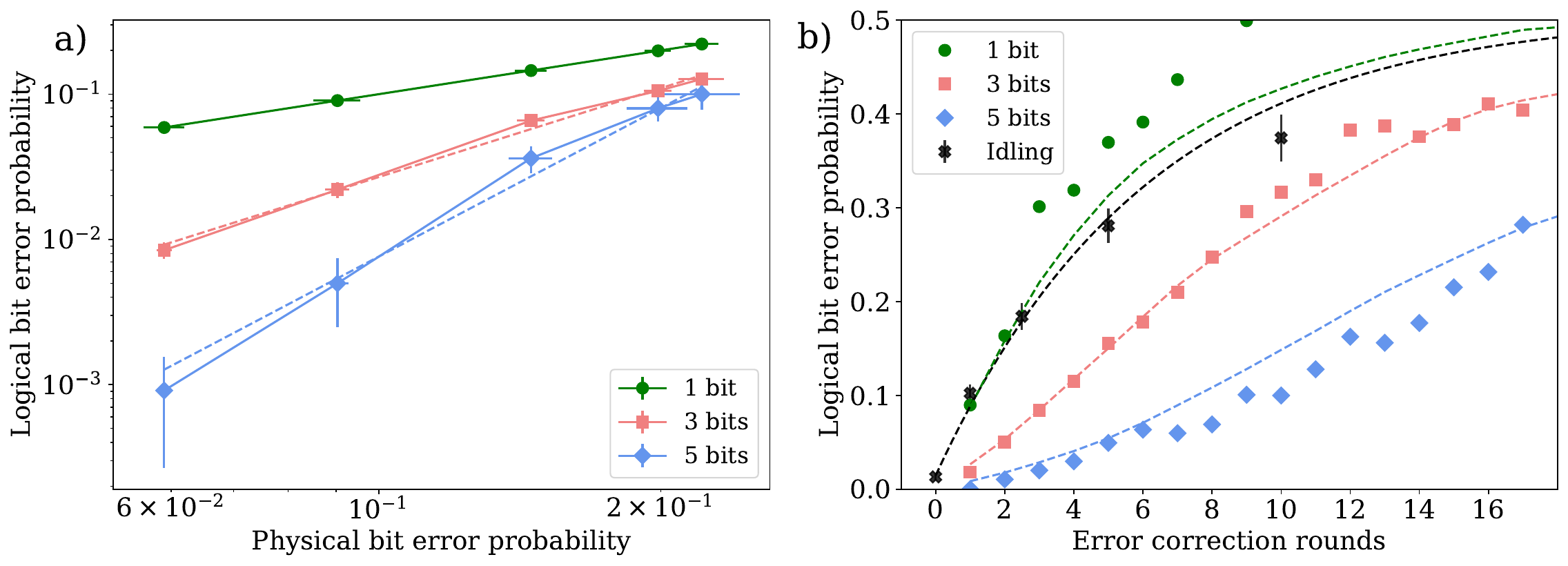}
\caption{a) Exponential suppression of the logical error with increasing code distance. 
b) Despite losses, encoding a logical bit in 3 or 5 atomic bits results in a decreased chance of an error and increased lifetime compared to an idling atomic bit.
Dashed lines are Monte-Carlo simulations using single atom error probability (9\%) and loss probability (3.7\%) extracted from the same dataset.}
\label{atomic_register}
\end{figure*}

As a preparation for future cavity-enabled quantum error correction in an array of atoms with Rydberg interactions \cite{Evered2023, Ma2023, tsai2024, Bluvstein2022, Bluvstein2024, radnaev2024},
we implement repeated rounds of classical error correction with a register of atomic bits. 
In each experiment, a logical bit $\{0,1\}$ is encoded in the hyperfine state $F=\{1, 2\}$ of multiple atoms. 
We introduce bit flip errors on each atom with some probability by varying the idling time between state preparation and measurement, with idling errors dominated by background depumping from \SI{808}{nm} tweezers.
All atoms are measured sequentially, and a majority vote is taken to determine the logical state. 
If the majority vote results in a tie, or if no atoms remain, the result is given by a coin toss. 
Atoms are then re-initialized according to the measurement result, and this cycle of preparation, measurement and error correction continues for up to 17 rounds.

In Fig.\:\ref{atomic_register}a, we post-select on the number of atoms present in each error-correction round, and extract logical bit error probability for distance $d=1$, $3$ and $5$ logical bits given varying physical bit error probabilities.
For one atomic bit, the logical error probability equals the physical error probability.
We fit the $d=3$ and $5$ data to power laws, finding exponents $2.0(1)$ and $3.4(3)$, respectively. 
The fitted exponents are consistent with $(d+1)/2$, in agreement with the expectation of exponential suppression of logical bit error rate with increasing distance $d$. 

In an error correction scenario with atom loss, measurement-induced or vacuum losses shrink the code distance throughout repeated rounds. 
Without replenishing the lost atoms, we nonetheless observe in Fig.\:\ref{atomic_register}b that losses are sufficiently low for a logical bit encoded in $3$ or $5$ physical bits to be preserved beyond the physical-bit idling lifetime. 
The physical-bit idling lifetime, \SI{125}{ms}, is a combination of separately measured depump/repump timescale (\SI{150}{ms}) and vacuum lifetime (\SI{800}{ms}). 
Varying the idling time between rounds, we find that the logical lifetime is maximized with \SI{20}{ms} of idling, where measurement loss and vacuum loss are balanced. 
A $d=1$ logical bit has slightly worse performance than a physical idling bit due to measurement-induced losses and errors. 
For $d=3$ and $5$ logical bits, we observe a significant reduction in the logical bit error probability, and the logical bit lifetime (defined as the time when error probability reaches $1/e$ of its final value) is extended by a factor of 2.5 and 4.9 compared to the physical-bit idling lifetime.

Although scaling to larger codes would further extend the logical memory, even percent-level losses would accumulate rapidly, degrading the code distance and ultimately limiting the logical lifetime. 
This highlights the central importance of replenishing lost atoms.
Non-destructive readout reduces but cannot eliminate the necessity of replenishing atoms when performing many rounds of error correction and deep circuits.
In our experiment, for example, perpetual operation of a $d=5$ logical bit would require replenishing atoms at a relaxed rate of \SI{10}{Hz}, well below what has been experimentally demonstrated \cite{gyger2024, Norcia2024}. 

In summary, we have realized for the first time site-selective hyperfine-state readout of a 10-site $^{87}$Rb atom array with an optical cavity.
For application to quantum error correction, we have demonstrated the use of adaptive searches to further speed up syndrome readout.
In the future, our site-selective cavity readout can be applied with microsecond switching time to any subset of a 1D or 2D array of hundreds of atoms with commercial 
 lasers.
For even larger arrays, we envision using the cavity mode as a readout zone, with batches of atoms transported in and out.
Combined with state-of-the-art cavities \cite{Gehr2010, Deist2022} capable of $\SI{10}{\mu s}$ readout times, our site-selective adaptive approach could enable non-destructive readout of 1000 syndrome qubits containing 1\% errors in under a millisecond.
Finally, we have demonstrated 17 rounds of real-time classical error correction of a logical bit encoded in an atomic register, marking the first classical repetitive error correction in neutral atoms.
When combined with Rydberg quantum gates \cite{Evered2023,Ma2023, tsai2024, radnaev2024} and single-qubit rotations \cite{Levine2022}, the demonstrated approach will enable repeated rounds of full quantum error correction.
For different applications, the site-selective excited-state shift can be used to control atom-cavity coupling for multiplexed entanglement generation \cite{Huie2021, Covey2023, Hartung2024, Krutyanskiy2023b, Krutyanskiy2024, Cetina2013, sinclair2024} or programmable atom-atom interactions \cite{Welte2018, Vaidya2018, Periwal2021, Ramette2022_any, Sauerwein2023}, and may also prove useful for continuous reloading and mid-circuit readout in free space.

We gratefully acknowledge Tout Wang and Mikhail D. Lukin for valuable discussions as well as loaning us the \SI{1529}{nm} laser, Luke Stewart for contributions to the experimental setup, and Ziv Aqua for comments on the manuscript. 
This project was funded in part by DARPA under the ONISQ program (grant \# 134371-5113608), the MIT-Harvard Center for Ultracold Atoms (NSF grant \# PHY-1734011), Quera, and the ARO (grant \# W911NF1910517). Support is also acknowledged from the U.S. Department of Energy, Office of Science, National Quantum Information Science Research Centers, Quantum Systems Accelerator (contract \# 7571809).

\bibliography{main}

\end{document}